\documentclass[aps,floatfix,preprint,preprintnumbers,nofootinbib,superscriptaddress,natbib]{revtex4}

\usepackage{graphicx,float}% Include figure files
\usepackage{epsfig}		
\usepackage{dcolumn}% Align table columns on decimal point
\usepackage{bm}% bold math
\usepackage{subfigure}

\usepackage[all]{xy}
\usepackage{amsmath,upgreek}
\usepackage{amssymb}

\usepackage{pdfpages}
\usepackage{color}
\usepackage{graphicx,epstopdf}
\usepackage[colorlinks=true,
 linkcolor=red,
 urlcolor=purple,
 citecolor=blue,hyperindex]{hyperref}
 \hypersetup{%
 colorlinks=true, 
 citecolor=blue,
 linkcolor=blue,
 urlcolor=black,
 filecolor=magenta
 }
\def\0{\mbox{\tiny $0$}}
\def\1{\mbox{\tiny $1$}}
\def\2{\mbox{\tiny $2$}}
\def\3{\mbox{\tiny $3$}}
\def\4{\mbox{\tiny $4$}}
\def\5{\mbox{\tiny $5$}}
\def\6{\mbox{\tiny $6$}}
\def\7{\mbox{\tiny $7$}}
\def\8{\mbox{\tiny $8$}}
\def\9{\mbox{\tiny $9$}}

\def\f14{\mbox{\tiny $\frac{1}{4}$}}

\def\ftn#1{{\color{purple}\footnote{\color{purple}{#1}}}}
\begin{document}

\title{Emergent time crystals from phase-space noncommutative quantum mechanics}

\author{A. E. Bernardini}
\email{alexeb@ufscar.br}
\altaffiliation[On leave of absence from]{~Departamento de F\'{\i}sica, Universidade Federal de S\~ao Carlos, PO Box 676, 13565-905, S\~ao Carlos, SP, Brasil.}
\author{O. Bertolami}
\email{orfeu.bertolami@fc.up.pt}
\altaffiliation[Also at~]{Centro de F\'isica das Universidades do Minho e do Porto, Rua do Campo Alegre s/n, 4169-007, Porto, Portugal.} 
\affiliation{Departamento de F\'isica e Astronomia, Faculdade de Ci\^{e}ncias da
Universidade do Porto, Rua do Campo Alegre 687, 4169-007, Porto, Portugal.}
\date{\today}% It is always \today, today,
 % but any date may be explicitly specified

\begin{abstract}
It has been argued that the existence of time crystals requires a spontaneous breakdown of the continuous time translation symmetry so to account for the unexpected non-stationary behavior of quantum observables in the ground state.
Our point is that such effects do emerge from position ($\hat{q}_i$) and/or momentum ($\hat{p}_i$) noncommutativity, i.e., from $[\hat{q}_i,\,\hat{q}_j]\neq 0$ and/or $[\hat{p}_i,\,\hat{p}_j]\neq 0$ (for $i\neq j$).
In such a context, a predictive analysis is carried out for the $2$-dim noncommutative quantum harmonic oscillator through a procedure supported by the Weyl-Wigner-Groenewold-Moyal framework.
This allows for the understanding of how the phase-space noncommutativity drives the amplitude of periodic oscillations identified as time crystals.
A natural extension of our analysis also shows how the spontaneous formation of time quasi-crystals can arise.
\end{abstract}
%\pacs{03.65.-w, 03.67.-a, }
\date{\today}
\maketitle

Time crystals are time-periodic self-organized structures that are supposed to emerge in the time domain due to the spontaneous breaking of time translation symmetry \cite{W1,W2}. They are analogous to spatial crystal lattices that are formed when the spontaneous breaking of space translation symmetry takes place \cite{PR2018}. 
Their features have been quantum mechanically probed in ultra-cold atoms \cite{Sacha15,Smits18} and spin-based solid state \cite{Khemani16,Else16,Pal,Rovny,Kyprianidis,Randall,Mi} systems, through which it has been claimed that periodically driven systems exhibit a discrete time symmetry which corresponds to a time translation led by the period of an external driver \cite{Zhang,Choi}.
Such plethora of follow up experiments \cite{Sacha15,Smits18,Khemani16,Else16,Pal,Mi,Zhang,Choi} suggests the emergence of novel phases of matter \cite{Rovny,Kyprianidis,Randall} that exhibit a discrete time translation symmetry, mostly described as arising from the breakdown of the continuous time translation symmetry, $\hat{\mathcal{T}}_H\equiv e^{-i\hat{H}t}$.
 
In fact, if a time-independent system driven by a time-independent Hamiltonian $H$ is prepared in an eigenstate $\vert\psi_n\rangle$, such that $H \vert\psi_n\rangle = E_n \vert\psi_n\rangle$, for the energy eigenvalue $E_n$, quantum mechanics (QM) implies that the probability density at a fixed position in the configuration space is also time-independent. Nevertheless, according to the above arguments and experiments, the existence of time crystals would admit that $[\hat{H},\,\rho_n]\equiv[\hat{\mathcal{T}}_H,\,\rho_n]\neq 0$ for $\rho_n = \vert\psi_n\rangle\langle\psi_n\vert$, which would correspond to a spontaneous breakdown of time translation symmetry followed by a non-stationary behavior of the eigensystem solutions.

In this letter, we argue that such a non-stationary behavior, and its straightforward connection with time crystal properties, both emerge from position and/or momentum noncommutativity in the phase-space.

As its inception, noncommutativity was firstly considered in the space coordinate domain as a way to regularize quantum field theories \cite{Snyder47}. Subsequently it appeared in the formulation of string theories \cite{Connes,Douglas,Seiberg,Nekrasov01}.
As a natural extension, the phase-space noncommutative (NC) QM considered here \cite{Gamboa,06A,Rosenbaum,Bastos,08A,09A} was formulated in terms of the Weyl-Wigner-Groenewold-Moyal (WWGM) framework \cite{Groenewold,Moyal,Wigner}, which is supported by a $2n$-dim phase-space that satisfies a deformed Heisenberg-Weyl algebr, where position and momentum operators,  $\hat{q}_i$ and $\hat{p}_j$, obey the following commutation relations,
\begin{equation}
[ \hat{q}_i, \hat{q}_j ] = i \theta_{ij} , \hspace{0.5 cm} [ \hat{q}_i, \hat{p}_j ] = i \hbar \delta_{ij} ,
\hspace{0.5 cm} [ \hat{p}_i, \hat{p}_j ] = i \eta_{ij},\label{EQEq31}
\end{equation}
with $ i,j= 1, ... ,d$, and $\eta_{ij}$ and $\theta_{ij}$ identified as entries of invertible antisymmetric real constant ($d \times d$) matrices, $ {\bf \Theta}$ and ${\bf N}$, such that an equally invertible matrix, ${\bf \Sigma}$, with $\Sigma_{ij} \equiv \delta_{ij} + \hbar^{-2} \theta_{ik} \eta_{kj}$, is identified if $\theta_{ik}\eta_{kj} \neq -\hbar^2 \delta_{ij}$.
Of course, given that $\eta_{ij},\,\theta_{ij} \neq 0$, the relations from Eq.~\eqref{EQEq31} can affect the symmetries related to conserved quantities usually identified by generic QM operators, $\hat{\mathcal{O}}$, for which
$d \langle\hat{\mathcal{O}}\rangle/dt = i\hbar^{-1}\,\langle[\hat{H},\,\hat{\mathcal{O}}]\rangle =0$.

Thus, for quantum operators identified by $\hat{\mathcal{O}} \to \hat{\mathcal{O}}(\{\hat{q}_i,\hat{p}_i\})$, the central question examined here is whether, in such NC extension of QM, the time crystal behavior emerges from the breakdown of time translational symmetries identified by $\langle[\hat{H},\,\hat{\mathcal{O}}(\{\hat{q}_i,\hat{p}_i\})]\rangle \neq 0$, in contrast to the standard QM prediction, $\langle[\hat{H},\,\hat{\mathcal{O}}]\rangle =\langle[\hat{H},\,\hat{\mathcal{O}}(\{\hat{q}_i,\hat{p}_i\})]\rangle{\vert}_{\eta_{ij}=\theta_{ij}=0} = 0$.
To investigate this point, the NC algebra from Eq.~\eqref{EQEq31} can be mapped into the Heisenberg-Weyl algebra through the linear Seiberg-Witten (SW) transformation \cite{Seiberg}, 
\begin{equation}
 \hat{q}_i = A_{ij} \mathit{\hat{Q}}_j + B_{ij} \hat{\Pi}_j \hspace{1 cm}
 \hat{p}_i = C_{ij} \mathit{\hat{Q}}_j + D_{ij} \hat{\Pi}_j,
\label{EQEq35}
\end{equation}
where $A_{ij}, B_{ij}, C_{ij}$ and $D_{ij}$ are real entries of constant matrices, ${\bf A}, {\bf B}, {\bf C}$ and ${\bf D}$. 
In this case, one recovers the algebra of ordinary QM,
\begin{equation}
[ \hat{Q}_i, \hat{Q}_j ] = 0, \quad [ \hat{Q}_i, \hat{P}_j ]
= i \hbar \delta_{ij} , \hspace{0.5 cm} [ \hat{P}_i, \hat{P}_j ] = 0,
\label{EQNCeq}
\end{equation}
from which it is straightforward to obtain the following matrix equation constraints \cite{Rosenbaum},
${\bf A} {\bf D}^T - {\bf B} {\bf C}^T = {\bf I}_{d \times d}$, ${\bf A} {\bf B}^T - {\bf B} {\bf A}^T = \hbar^{-1} {\bf \Theta}$, and ${\bf C} {\bf D}^T - {\bf D} {\bf C}^T = {\hbar^{-1}} {\bf N}$,
where the superscript $T$ denotes matrix transposition.

The algebra, Eq.~\eqref{EQEq31}, in the context of the WWGM framework \cite{Bastos,Rosenbaum}, allows for examining striking features which include putative violations of the Robertson-Schr\"odinger uncertainty relation \cite{Bastos001,Bastos002}, quantum correlations and information collapse in gaussian quantum systems \cite{Bernardini13B,Bernardini13B2,RSUP1,RSUP2,RSUP3}, and regularizing features in minisuperspace quantum cosmology models \cite{Bastos001,Bastos003} and in black-hole physics \cite{Bastos004,Bastos005}.
The generalized WWGM {\em star}-product, the extended Moyal bracket and the NC Wigner function framework ensure that observables are independent of any particular choice of the SW map \cite{Bastos}.

To make our proposal more concrete, the $2$-dim harmonic oscillator in the NC phase-space \cite{Bernardini13A} will be evaluated in order to show that time crystal patterns on quantum observables and quantum states naturally emerge from the NC QM. Hence, let us consider the quantum Hamiltonian,
\begin{equation}\label{EQNCCC}
\hat{H}_{HO}(\hat{\mathbf{q}},\hat{\mathbf{p}}) = \frac{\hat{\mathbf{p}}^2}{2m} + \frac{1}{2}m \omega^2 \hat{\mathbf{q}}^2,
\end{equation}
on the NC ``$x-y$'' plane, with position and momentum satisfying the NC algebra, Eq.~\eqref{EQEq31}, now with $i,j=1,2$, $\theta_{ij} = \theta \epsilon_{ij}$ and $\eta_{ij} = \eta \epsilon_{ij}$, where $\epsilon_{ij}$ is the $2$-dim Levi-Civita tensor. The map to commutative operators is given by
\begin{eqnarray}
\mathit{\hat{Q}}_i &=& \mu \left(1 - {\theta \eta\over\hbar^2} \right)^{- 1 / 2} \left( \hat{q}_i + {\theta\over 2 \lambda \mu \hbar} \epsilon_{ij} \hat{p}_j \right)~,\nonumber\\
\hat{\Pi}_i &=& \lambda \left(1 - {\theta \eta\over\hbar^2} \right)^{-1 / 2} \left( \hat{p}_i-{\eta\over 2 \lambda \mu \hbar} \epsilon_{ij} \hat{q}_j \right),
\label{EQSWinverse}
\end{eqnarray}
in terms of the SW map,
%\ftn{For which the corresponding Jacobian reads
%\begin{equation}
%{\partial (q,p)\over \partial (\mathit{Q}, \Pi)} = (\det {\mathbf \Omega})^{1/2}=1 - {\theta \eta\over\hbar^2}.
%\end{equation}},
\begin{equation}
 \hat{q}_i = \lambda \mathit{\hat{Q}}_i - {\theta\over2 \lambda \hbar} \epsilon_{ij} \hat{\Pi}_j \hspace{0.5 cm},\hspace{0.5 cm} \hat{p}_i = \mu \hat{\Pi}_i + {\eta\over 2 \mu \hbar} \epsilon_{ij} \mathit{\hat{Q}}_j~,
\label{EQSWmap}
\end{equation}
which is invertible for $\theta\eta \neq \hbar^2$, and the parameters $\lambda$ and $\mu$ satisfying the condition
\begin{equation}
{\theta \eta \over 4 \hbar^2} = \lambda \mu ( 1 - \lambda \mu ).%\qquad (\theta\eta \lesssim \hbar^2).
\label{EQconstraint}
\end{equation}

\begin{figure}[h!]
\includegraphics[width= 9 cm]{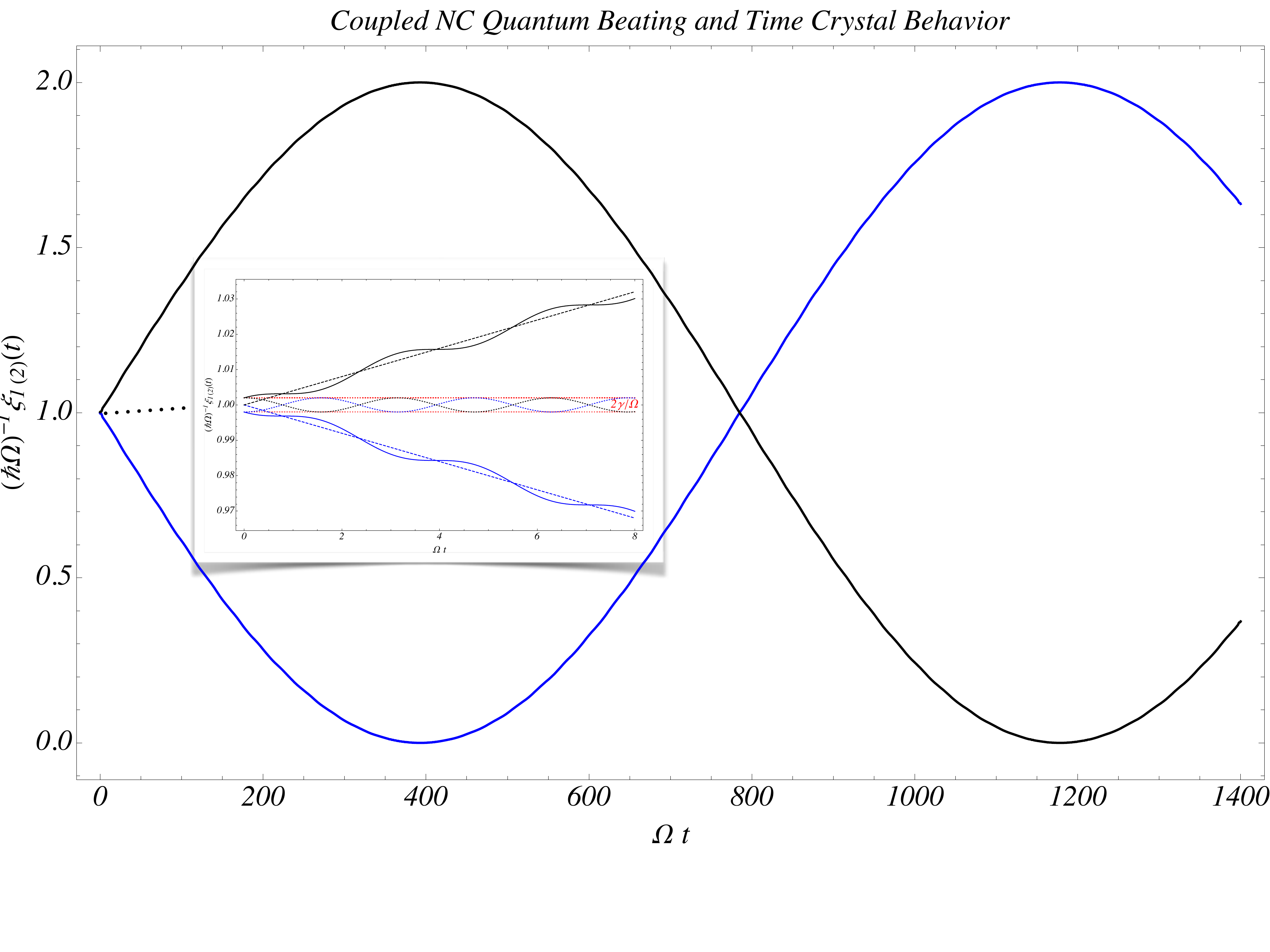}
\vspace{-1. cm}\renewcommand{\baselinestretch}{1}
\caption{(Color online) Dimensionless NC associated energies , $(\hbar\Omega)^{-1}{\xi}_{1(2)}$ (black (blue) line) as function of $\Omega t$, for $\gamma /\Omega=0.002$. Decoupled $\gamma$-frequency NC quantum beating (dashed lines) and $\Omega$-frequency time crystal behavior (dotted lines) for $\gamma t \gtrsim 0$ are identified in the {\em zoom in} window. 
}
\label{EQTC0004}
\end{figure}

\begin{figure}[h!!]
\includegraphics[width= 9 cm]{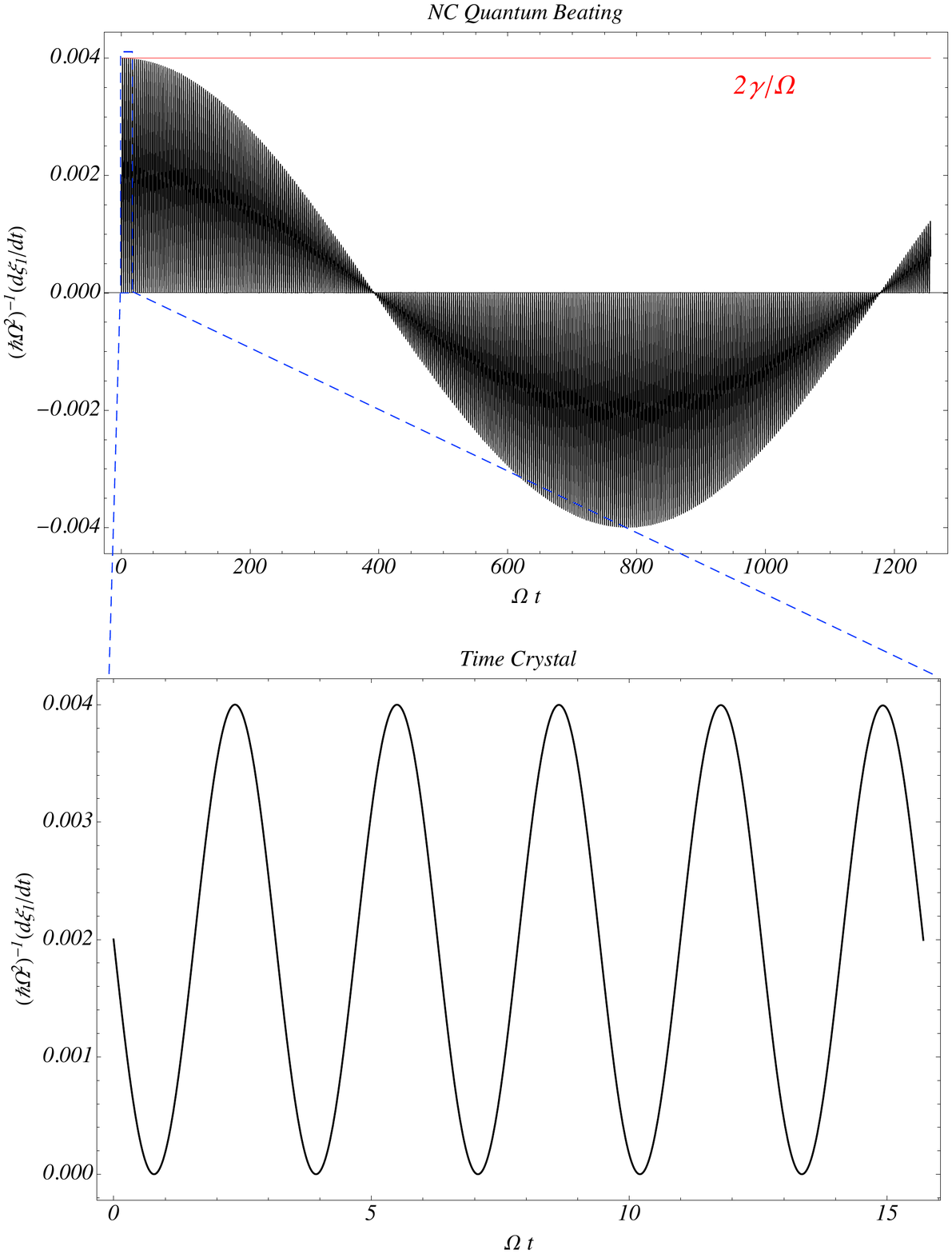}
\vspace{-1. cm}\renewcommand{\baselinestretch}{1}\caption{(Color online)Dimensionless time derivative, $(\hbar\Omega^2)^{-1}\dot{\xi}_1$ as function of $\Omega t$, for $\gamma /\Omega=0.002$. The NC beating behavior is depicted by the first plot and the time crystal periodic behavior, driven by the amplitude modulation, $\hbar \gamma \Omega(\equiv \gamma/\Omega \times \hbar \Omega^2)$ (red line) is depicted by the second (zoom in) plot.}
\label{EQTC0001}
\end{figure}

The Hamiltonian in terms of the so-called commutative variables, $\mathit{\hat{Q}}_i$ and $\hat{\Pi}_i$, reads \cite{Bernardini13A}
\begin{equation}
\hat{H}_{HO}(\hat{\mbox{\bf \em Q}},\hat{\mathbf{\Pi}}) = \alpha^2\hat{\mbox{\bf \em Q}}^2 +\beta^2\hat{\mathbf{\Pi}}^2 + \gamma \sum_{i,j = 1}^2{\epsilon_{ij}\hat{\Pi}_i \mathit{\hat{Q}}_j},
\label{EQHamilton}
\end{equation}
where  ${\alpha}^2 \equiv m  \omega^2\lambda^2/2 + \eta^2/(8m  \hbar^2\mu^2)$, ${\beta}^2 \equiv {\mu^2/(2m)} + {m \omega^2 \theta^2/(8 \hbar^2 \lambda^2)}$, and ${\gamma} \equiv m \omega^2{\theta}/({2\hbar}) + {\eta}/({2m\hbar})$, from which one obtains the following set of coupled equations of motion,
\begin{eqnarray}
\dot{\Pi}_i &=& -\frac{i}{\hbar} \langle\left[\hat{\Pi}_i,\,\hat{H}_{HO}\right]\rangle = -2 \alpha^2\,\mathit{Q}_i - \gamma\,\varepsilon_{ji}\Pi_j,\nonumber\\
\dot{\mathit{Q}}_i &=& -\frac{i}{\hbar} \langle\left[\hat{\mathit{Q}}_i,\,\hat{H}_{HO}\right]\rangle = ~~2 \beta^2\,\Pi_i - \gamma\,\varepsilon_{ji}\mathit{Q}_j,
\label{EQeqs01}
\end{eqnarray}
with $\mathit{Q}_i \equiv \langle \hat{\mathit{Q}}_i \rangle$ and ${\Pi}_i \equiv \langle \hat{\Pi}_i \rangle$. In this case, $\mbox{\bf \em Q} = (\mathit{Q}_1,\,\mathit{Q}_2)$ and $\mathbf{\Pi}= ({\Pi}_1,\,{\Pi}_2)$ may be interpreted as the dynamical variables within the WWGM formalism for which the solutions are given by \cite{Bernardini13A}
\small
\begin{eqnarray}
\mathit{Q}_1(t)&=& x\,\cos(\Omega t)\cos(\gamma t) + y\,\cos(\Omega t)\sin(\gamma t)
 +\frac{\beta}{\alpha}\left[\pi_y\,\sin(\Omega t)\sin(\gamma t) + \pi_x\,\sin(\Omega t)\cos(\gamma t)
\right],\nonumber\\
\mathit{Q}_2(t)&=& y\,\cos(\Omega t)\cos(\gamma t) - x\,\cos(\Omega t)\sin(\gamma t)
 -\frac{\beta}{\alpha}\left[\pi_x\,\sin(\Omega t)\sin(\gamma t) - \pi_y\,\sin(\Omega t)\cos(\gamma t)
\right],\nonumber\\
\Pi_1(t)&=& \pi_x\,\cos(\Omega t)\cos(\gamma t) + \pi_y\,\cos(\Omega t)\sin(\gamma t)
 -\frac{\alpha}{\beta}\left[y\,\sin(\Omega t)\sin(\gamma t) + x\,\sin(\Omega t)\cos(\gamma t)
\right],~~\nonumber\\
\Pi_2(t)&=& \pi_y\,\cos(\Omega t)\cos(\gamma t) - \pi_x\,\cos(\Omega t)\sin(\gamma t)
 +\frac{\alpha}{\beta}\left[x\,\sin(\Omega t)\sin(\gamma t) - y\,
\sin(\Omega t)\cos(\gamma t)
\right],~~
\label{EQsolutions}
\end{eqnarray}
\normalsize
where $x,\, y,\,\pi_x,$ and $\pi_y$ are arbitrary parameters, and 
\begin{equation}
\Omega = 2 \alpha \beta = \sqrt{(2\lambda\mu - 1)^2\omega^2 + \gamma^2} = \sqrt{\omega^2 +\gamma^2 - {\theta \eta \over \hbar^2}},
\label{EQeq37}
\end{equation}
with $\lambda$ and $\mu$ being eliminated by the constraint Eq.~\eqref{EQconstraint}.
In particular, by setting $\theta=\eta = 0$, and therefore $\gamma = 0$, one recovers the solutions for the $2$-dim harmonic oscillator with uncoupled $x-y$ coordinates and $\Omega = \omega$.
For $\theta,\, \eta \neq 0$, the above results lead to two decoupled time-invariant quantities,
\begin{eqnarray}
\sum_{i=1}^2{\left(\frac{\alpha}{\beta} \mathit{Q}_i(t)^2 + \frac{\beta}{\alpha}\Pi_i(t)^2\right)} &=& \frac{\alpha}{\beta} (x^2 + y^2) + \frac{\beta}{\alpha}(\pi_x^2 + \pi_y^2),\nonumber\\
\sum_{i,j=1}^2{\left(\epsilon_{ij} \mathit{Q}_i(t)\,\Pi_j(t)\right)} &=& x\,\pi_y - y \,\pi_x.
\label{EQinvariant}
\end{eqnarray}
The meaning of the modifications introduced by the NC variables can be evinced by setting $\pi_x = \pi_y = \sqrt{\alpha\hbar/2\beta}$, and $x = y = \sqrt{\beta\hbar/2\alpha}$, so that the associated $x$ and $y$ translational energy contributions can be shown to evolve as 
\begin{equation}\small
E_i =\alpha\beta {\left(\frac{\alpha}{\beta} \mathit{Q}_i(t)^2 + \frac{\beta}{\alpha}\Pi_i(t)^2\right)} = \frac{\hbar\Omega}{2} \left(1 - (-1)^i\,\sin(2\gamma t)\right),
\label{EQenergy}
\end{equation}\normalsize
with $i = 1,\,2$, from which a typical low frequency $\gamma$-dependent beating behavior is found \cite{Bernardini13A}.
Such a time-dependent periodic modification is a clear new feature of the NC harmonic oscillator ground state associated energy\ftn{The {\em stargen}functions for the Hamiltonian, Eq.~(\ref{EQHamilton}), are obtained from the {\em stargen}value equation,
\begin{equation}
H^W_{HO} \star \rho_{n_{\tiny 1},n_{\tiny 2}}^W (\mbox{\bf \em Q},\mathbf{\Pi}) = 
E_{n_{\tiny{1}},n_{\tiny{2}}}\,\rho^W_{n_{\tiny{1}},n_{\tiny{2}}} (\mbox{\bf \em Q},\mathbf{\Pi}),
\label{EQhelp01}
\end{equation}
where $W (\mbox{\bf \em Q},\mathbf{\Pi})$ is the eigenstate associated Wigner function, from which one has \cite{Rosenbaum},
\begin{equation}
\rho_{n_{\tiny{1}},n_{\tiny{2}}}^W (\mbox{\bf \em Q},\mathbf{\Pi}) = \frac{(-1)^{n_1+n_2}}{\pi^2\hbar^{2}}\exp\left[{-\frac{1}{\hbar}\left(\frac{\alpha}{\beta}\mbox{\bf \em Q}^2 + \frac{\beta}{\alpha}{\mathbf{\Pi}}^2\right)}\right] \, L^0_{n_1} \left(\Omega_{+}/\hbar\right) \,L^0_{n_2}\left(\Omega_{-}/\hbar\right),
\label{EQLague01}
\end{equation}
where $L^0_n$ are the associated Laguerre polynomials, $n_1$ and $n_2$ are non-negative integers, and
\begin{equation}
{\Omega}_{\pm} = {\alpha\over\beta}\mbox{\bf \em Q}^2 + {\beta\over\alpha}\mathbf{\Pi}^2 \mp 2 \sum_{i,j = 1}^2{\left(\epsilon_{ij}\Pi_i \mathit{Q}_j\right)},
\end{equation}
such that the energy spectrum is given by
$E_{n_{\tiny 1},n_{\tiny 2}} = \hbar\left[2\alpha\beta(n_1 + n_2 + 1) + \gamma (n_1 - n_2)\right]$.}. In order to clarify the relation with the time crystal behavior, one should get back to the Hamiltonian Eq.~\eqref{EQNCCC} and examine the contributions of $\{\hat{q}_1,\,\hat{p}_1\}$ and $\{\hat{q}_2,\,\hat{p}_2\}$ to the energy and eigenstates. 
By identifying the associated energy of each $i$-sector ($i=1,\,2$)  as
\begin{equation}\label{EQNCCC2}
\hat{\xi}_i = \frac{\hat{p}_i^2}{2m} + \frac{1}{2}m \omega^2 \hat{q}_i^2,
\end{equation}
from standard QM, one should have $\dot{\xi}_i = i\hbar^{-1} \langle [H,\, \hat{\xi}_i]\rangle = 0$ (with $\langle \hat{\xi}_i\rangle \equiv {\xi}_i$).
However, after recasting $\{\hat{q}_i,\,\hat{p}_i\}$ in terms of the SW map, Eqs.~\eqref{EQSWinverse}-\eqref{EQSWmap}, with $\Omega$, $\omega$ and $\gamma$ constrained by Eq.~\eqref{EQeq37}, one obtains an unexpected non-stationary behavior for each of the energy contributions,
\begin{eqnarray}\label{EQNCCC3}
\xi_i(t) &=& 
\frac{\hbar \Omega}{2}
\left\{1 - (-1)^i
\left[
\sqrt{1-\frac{\omega ^2}{\Omega ^2}}\left(\cos (2 \gamma t)\, \cos (2 \Omega t )-
\frac{\gamma}{\Omega} \sin (2 \gamma t)\, \sin (2 \Omega t )\right)\right.\right.\\
&&\qquad\qquad\qquad\qquad\qquad\qquad\qquad\qquad\qquad\qquad\qquad\left.\left.+\frac{\omega}{\Omega}\sqrt{1-\frac{\gamma ^2}{\Omega ^2}}\sin (2 \gamma t)
\right]\right\},\nonumber
\end{eqnarray}
from which the time crystal non-stationary behavior driven by $\Omega$ and a beating behavior driven by $\gamma$ both emerge. As depicted in Fig.~\ref{EQTC0004}, if either $\theta$ or $\eta$ vanishes, one has $\Omega^2 = \omega^2 +\gamma^2$ and
\small\begin{equation}
\xi_i(t)=
\frac{\hbar \Omega}{2}
\left\{1 - (-1)^i
\left[
\frac{\gamma}{\Omega}\left(\cos (2 \gamma t)\, \cos (2 \Omega t )-
\frac{\gamma}{\Omega} \sin (2 \gamma t)\, \sin (2 \Omega t )\right)+\left(1-\frac{\gamma ^2}{\Omega ^2}\right)\sin (2 \gamma t)
\right]\right\}.
%\frac{d}{dt} \langle\hat{\xi}_i\rangle &=& 
%(-1)^i\hbar 
%\left\{\gamma\Omega^{-1}\left[\left(\gamma ^2-\Omega ^2\right) +\left(\gamma ^2+\Omega ^2\right) \sin (2 \Omega t )\right]\cos (2 \gamma t)+ 2 \gamma^2 \cos (2 \Omega t ) \sin (2 \gamma t) \right\}
\end{equation}\normalsize
For the arbitrary choice of $\gamma /\Omega=0.002$, in the smaller window of Fig.~\ref{EQTC0004}, the energy decoupled $\gamma$-frequency NC quantum beating (dashed lines) and the externally driven $\Omega$-frequency time crystal behavior (dotted lines) for $\gamma t \gtrsim 0$ can be clearly identified.
Correspondently, Fig.~\ref{EQTC0001} depicts the time derivative of the energy from which the magnitude of the time crystal oscillating behavior can be quantified. From Eq.~\eqref{EQNCCC3}, the externally driven oscillation amplitude, $\hbar \Omega/2$, is modulated by a factor $ \gamma/\Omega$.

Due to the presumed small magnitude of $\gamma$, the beating oscillations are probably difficult to measure. Conversely, an effect can be straightforwardly identified for $\gamma \ll \Omega$ and $\gamma t \gtrsim 0$ implying that 
\begin{equation}\label{EQNCCC4}
\xi_i(t)\approx 
\frac{\hbar \Omega}{2}\left[1 - (-1)^i \frac{\gamma}{\Omega} \left( 2\, \Omega t+\cos (2 \Omega t )\right)\right],
\end{equation}
at first order in $\gamma$.
In this case,
\begin{equation}\label{EQNCCC5}
\dot{\xi}_i(t)\approx 
(-1)^{i+1}\hbar \gamma\Omega\left[1- \sin (2 \Omega t)\right] ,
\end{equation}
a time crystal periodic behavior with a measurable energy time derivative oscillation amplitude, $\hbar \gamma \Omega$, driven by both the NC parameter, $\gamma$, and the external oscillation frequency $\Omega \sim \omega$.

\begin{figure}[b!]
\vspace{-1. cm}\includegraphics[width= 9.5 cm]{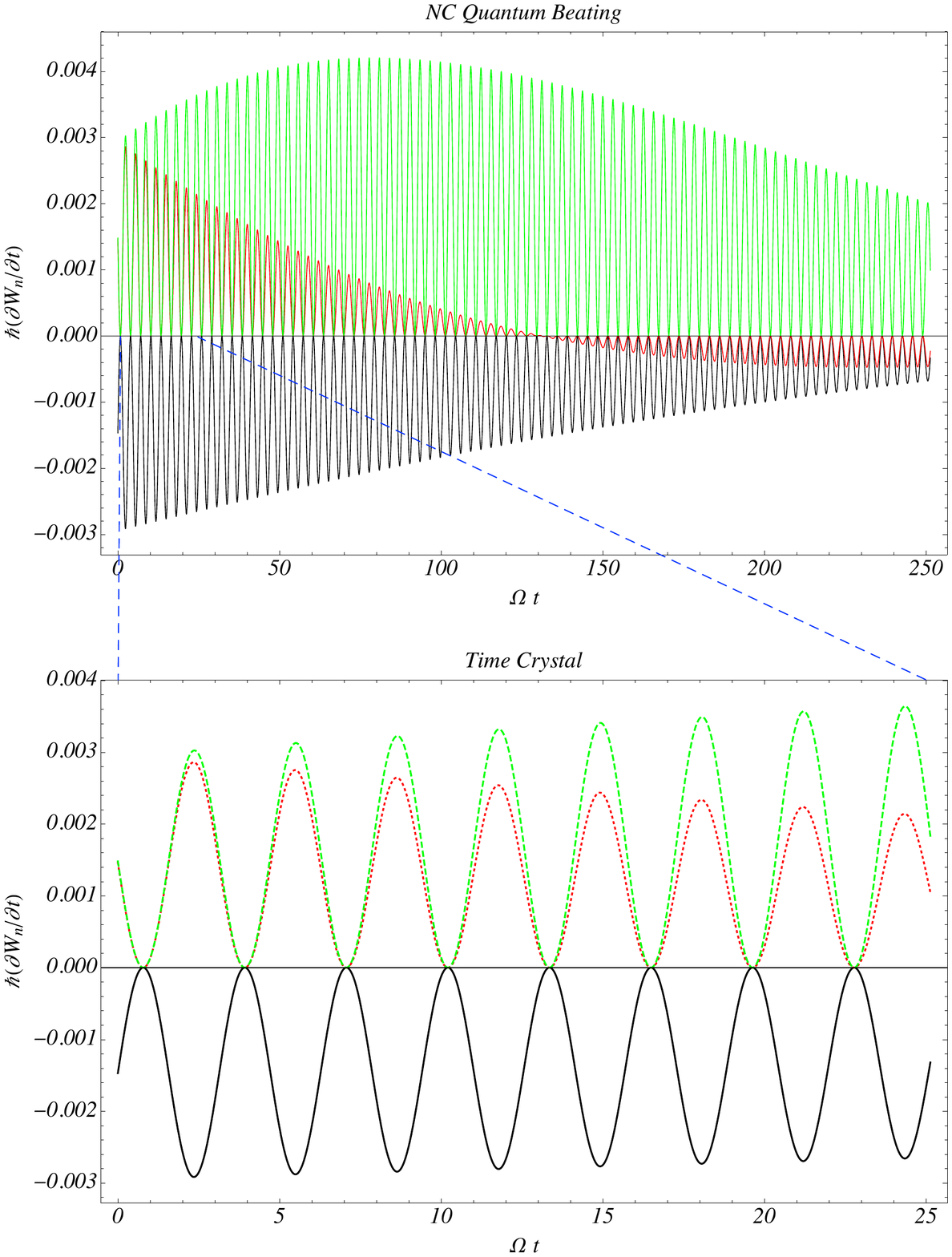}
\vspace{-1.5 cm}\caption{(Color online) Time dependent behavior of $\hbar\,\partial W_{n_{\tiny{1}}}(\xi_{1}(t)) /\partial t$ as function of $\Omega t$, for $\gamma /\Omega=0.002$, denoting the emergence of non-stationary behavior which is inexistent in standard QM. Plots are for ground state ($n_1= 0$, black solid lines), first ($n_1= 1$, red dotted lines) and second ($n_1= 2$, green dashed lines) excited states. Similar results can be obtained for $\partial W_{n_{\tiny{2}}}(\xi_{2}(t)) /\partial t$.
% in one (first plot) and two (second plot) dimensional representations, for $n_{\tiny{1}} = $n_{\tiny{1}} = 0,\, 1$ and $2$ and $$n_{\tiny{2}} = 0$
}
\label{EQTC0002}
\end{figure}

The same effects can be encountered on the behavior of the associated Wigner eigenfunctions (time derivatives) for each $i$-sector of the $2$-dim harmonic oscillator, as depicted in Fig.~\ref{EQTC0002}.
Once again, under the perspective of the standard QM, for which $H \, W_{n_{\tiny{1(2)}}}(\xi_{\tiny{1(2)}}(t)) =E _{n_{\tiny{1(2)}}} W_{n_{\tiny{1(2)}}}(\xi_{\tiny{1(2)}}(t))$, one would expect that the decoupled Wigner eigenfunctions, $W_{n_{\tiny{1}}}$ and $W_{n_{\tiny{2}}}$, written in terms of Laguerre polynomials, $L^0_{n}$,
\begin{equation}
\hbar\,W_{n_{\tiny{1(2)}}}(q_{\tiny{1(2)}},\,p_{\tiny{1(2)}}) \equiv \hbar\,W_{n_{\tiny{1(2)}}}(\xi_{\tiny{1(2)}}(t))
= \frac{1}{\pi} \exp{\left[-2\xi_{\tiny{1(2)}}(t)/\hbar\Omega\right]}
L^0_{n_{\tiny{1(2)}}} \left[4\xi_{\tiny{1(2)}}(t)/\hbar\Omega\right],
\end{equation}
should be stationary functions.

In fact, QM describes quantum superpositions of normalized solutions of uncoupled pairs of two 1-dim harmonic oscillators in terms of decoupled stationary functions, $\hbar\,W_{n_{\tiny{1}}}(\xi_{1}(t))$ and $\hbar\,W_{n_{\tiny{2}}}(\xi_{2}(t))$.
However, under the NC QM perspective, given the time dependent behavior of $\xi_{\tiny{1(2)}}(t)$, Fig.~\ref{EQTC0002} depicts the analogous non-stationary behavior in terms of, $\partial W_{n_{\tiny{1}}}(\xi_{1}(t)) /\partial t \neq 0$, for $n_{\tiny{1}} = 0,\, 1$ and $2$, from which the highly oscillatory behavior is also evinced.
Of course, for too small values of $\gamma /\Omega$ (say $\gamma /\Omega \ll 0.002$), time crystal and NC beating patterns would be hard to measure.

Interestingly, 2-dim (or 3-dim) Bose-Einstein condensates with time crystal behavior have, in fact, been detected through a resonance between two (or three) oscillating atom mirrors through a configuration where the localized wave-packets are products of 1-dim wave-packets \cite{Sacha0,Sacha1}. The results discussed above suggest a natural explanation for such effect, in this case, with {\em quasi}-periodic eigenstates driven by $\xi_1$, $\xi_2$, and $\gamma$ replacing the 1-dim wave-packets. In particular, for extended time intervals, in which the NC quantum beating takes place, the short time scale $\Omega$-frequency periodic behavior turns into a {\em quasi}-periodic one, due to the periodic corrections from $\gamma$-frequency driven contributions (cf. the Supplementary Material). 
This suggests that a putative connection of the above results with the spontaneous formation of time quasi-crystals from atoms bouncing between a pair of orthogonal atom mirrors \cite{Sacha0,Sacha2} also deserves further investigations.

To summarize, our results show that the spontaneous breaking of time translation symmetry featured by time crystals and their associated non-stationary behavior, which cannot be accounted by standard QM, emerge from either position ($q$) or momentum ($p$) noncommutativity, i.e., from $[\hat{q}_i,\,\hat{q}_j]\neq i \theta \epsilon_{ij}$ and/or $[\hat{p}_i,\,\hat{p}_j]\neq i \eta \epsilon_{ij}$.
In such a context, one can conclude that the NC parameters naturally drive the amplitude of periodic oscillations thought as time crystals.
Conversely, besides explaining the emergence of such unexpected properties, the measurable oscillation amplitude $\propto \hbar \gamma \Omega$ ($\gamma/\Omega$ in the dimensionless form) driven by both the NC parameter, $\gamma$, and the external oscillation frequency $\Omega \sim \omega$, can be tested in order to set bounds to the NC parameters.
Even though that might require either a more complex experimental engineering or a more elaborate set of harmonic oscillator configurations, the ideas considered here might stimulate further discussions on such a fascinating phenomena as time crystals.

{\em Acknowledgments -- The work of AEB is supported by the Brazilian Agency FAPESP (Grant No. 2020/01976-5)}.

\pagebreak
\pagenumbering{roman}
\pagenumbering{arabic}

\section*{Supplementary Material -- Emergent time crystals from phase-space noncommutative quantum mechanics}
%\title{Supplementary Material -- Emergent time crystals from phase-space noncommutative quantum mechanics}

\author{A. E. Bernardini}
\email{alexeb@ufscar.br}
\altaffiliation[On leave of absence from]{~Departamento de F\'{\i}sica, Universidade Federal de S\~ao Carlos, PO Box 676, 13565-905, S\~ao Carlos, SP, Brasil.}
\author{O. Bertolami}
\email{orfeu.bertolami@fc.up.pt}
\altaffiliation[Also at~]{Instituto de Plasmas e Fus\~ao Nuclear, Instituto Superior T\'ecnico, Av. Rovisco Pais, 1, 1049-001, Lisboa.} 
\affiliation{Departamento de F\'isica e Astronomia, Faculdade de Ci\^{e}ncias da
Universidade do Porto, Rua do Campo Alegre 687, 4169-007, Porto, Portugal.}
\date{\today}

%\begin{abstract}
%This is to exemplify how both time crystal and time {\em quasi}-crystal patterns emerges simultaneously arise from NC QM.
%\end{abstract}

In order to exemplify how both time crystal and time {\em quasi}-crystal patterns emerges simultaneously arise from NC QM, one should turn back to the choice of $x,\, y,\,\pi_x,$ and $\pi_y$ arbitrary parameters and constrain them by $\pi_x = \pi_y = k \sqrt{\alpha\hbar/2\beta}$, and $x = y = s \sqrt{\beta\hbar/2\alpha}$, with $k$ and $s$ corresponding to dimensionless phase-space coordinates; thus one can observe the NC effects on the $s-k$ plane.

\begin{figure}[h!]
\includegraphics[width= 7 cm]{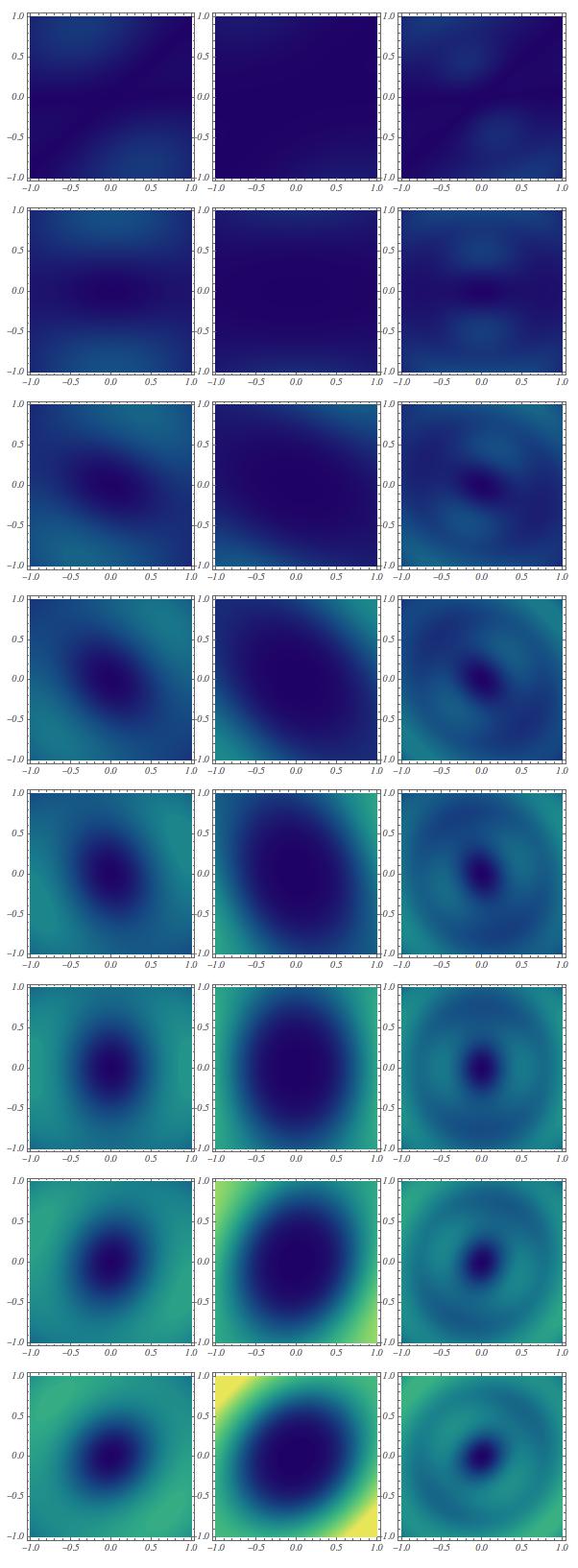}
\caption{\label{WignerF}\footnotesize{(Color Online) Phase-space (x-k) Wigner function described in terms of the difference between the NC Wigner function, $\pi^2W_{n_{\tiny{1}}}(\xi_{1}(t))\,W_{n_{\tiny{2}}}(\xi_{2}(t))$, and the standard QM Wigner function, $\pi^2 W_{n_{\tiny{1}}}((s^2+k^2)/2)\,W_{n_{\tiny{2}}}((s^2+k^2)/2)$, at $\Omega t = \ell\pi/8$ ($\ell^{th}$-row), with $\ell$ from $1$ to $8$. The {\em \
BlueGreenYellow} scale runs from 0 (blue) to 1 (yellow). Results are for $\gamma/\Omega = 0.002$, and $(n_1,\,n_2) = (0,\,1)$ (first column),$(1,\,1)$ (second column) and $(2,\,5)$ (third column). They are amplified by a factor $10^2$ in the first and third columns and by a factor $10^4$ in the second column.}}
\end{figure}

Fig.~\ref{WignerF}, depicts the phase-space $s-k$ Wigner function described in terms of the difference between the NC Wigner function, $\pi^2 W_{n_{\tiny{1}}}(\xi_{1}(t))\,W_{n_{\tiny{2}}}(\xi_{2}(t))$, and the standard QM Wigner function, $\pi^2 W_{n_{\tiny{1}}}((s^2+k^2)/2)\, W_{n_{\tiny{1}}}((s^2+k^2)/2)$, at $\Omega t = \ell\pi/8$ ($\ell^{th}$-row), with $\ell$ from $1$ to $8$.
The short time-scale pattern depicts the time crystal periodic behavior identified by an external
$\Omega$-frequency driven periodic rotation in the $s-k$ plane. A larger time-scale pattern (last row) shows that the time crystal periodic behavior is turned into a {\em quasi} periodic behavior identified by NC $\gamma$-dependent corrections, which suppresses the Wigner function amplitude. Hence, at a large time-scale, depending on the magnitude of the NC parameters, amplitude suppression is converted into amplitude revival, due to the NC $\gamma$-driven quantum beating. 

\end{document}